# Sub-cycle control of terahertz high-harmonic generation by dynamical Bloch oscillations

O. Schubert[1], M. Hohenleutner[1], F. Langer[1], B. Urbanek[1], C. Lange[1],
U. Huttner[2], D. Golde[2], T. Meier[3], M. Kira[2], S. W. Koch[2] and R. Huber[1]

[1] Department of Physics, University of Regensburg, 93040 Regensburg, Germany
[2] Department of Physics, University of Marburg, 35032 Marburg, Germany
[3] Department of Physics, University of Paderborn, 33098 Paderborn, Germany

**Ultrafast charge transport in strongly biased semiconductors is at the heart of high-speed electronics, electro-optics, and fundamental solid-state physics[1-13]. Intense light pulses in the terahertz (THz) spectral range have opened fascinating vistas[14-21]: Since THz photon energies are far below typical electronic interband resonances, a stable electromagnetic waveform may serve as a precisely adjustable bias[5,11,17,19]. Novel quantum phenomena have been anticipated for THz amplitudes reaching atomic field strengths[8-10]. We exploit controlled THz waveforms with peak fields of 72 MV/cm to drive coherent interband polarization combined with dynamical Bloch oscillations in semiconducting gallium selenide. These dynamics entail the emission of phase-stable high-harmonic transients, covering the entire THz-to-visible spectral domain between 0.1 and 675 THz. Quantum interference of different ionization paths of accelerated charge carriers is controlled via the waveform of the driving field and explained by a quantum theory of inter- and intraband dynamics. Our results pave the way towards all-coherent THz-rate electronics.**

Once called the 'THz gap', the spectral boundary between electronics and optics has recently been made accessible with unique photonics tools[19]. THz pulses with inherently stable carrier-envelope phase (CEP) and peak electric fields of 1 to 108 MV/cm have extended the scope of optics[14-16]. CEP-stable waveforms have facilitated resonant control of low-energy elementary excitations[18,19]. Alternatively these pulses may serve as ultrafast bias fields for extreme transport studies, since the photon energies lie far below electronic interband resonances of bulk semiconductors[17,19]. Acting on the femtosecond scale, THz biasing is not limited by d. c. dielectric breakdown. Peak fields of 1 MV/cm have shed new light on intervalley scattering or impact ionization[11]. For similar amplitudes an onset of ballistic transport through a fraction of the Brillouin zone of a bulk semiconductor has been reported[5].

Yet expectations are far ahead: A long-standing prediction by Bloch and Zener[1,2] states that the quasi-momentum $\hbar k$ of an electron in an electric field $E$ changes at a constant rate $\hbar dk/dt = -eE$, where $\hbar$ denotes the reduced Planck constant and $e$ is the elementary charge. If $E$ is large enough for $k$ to reach the edge of the Brillouin zone before scattering takes place, Bragg reflection occurs and the electron traverses the Brillouin zone over again. This scenario causes charge oscillations in real and reciprocal space. When the electric bias competes even with atomic potential gradients, extreme high-field phenomena such as interband ionization, Wannier-Stark localization and strong mixing of electronic bands may be expected. While analogous phenomena have been studied in artificial structures[22-27], the observation of coherent high-field transport in bulk solids has remained a challenge as the combination of ultrafast scattering4 and the small size of the crystalline unit cell requires extremely strong biasing. Here, we employ CEP-stable THz waveforms with amplitudes comparable to atomic fields to explore a qualitatively novel regime of ultrafast coherent charge transport. The field drives interband polarization in a bulk semiconductor off-resonantly and accelerates the excited electron-hole pairs to perform dynamical Bloch oscillations, generating CEP-stable radiation throughout the frequency range between 0.1 and 675 THz. Quantum interference of multiple excitation channels makes the dynamics highly sensitive to the CEP of the THz waveform.

Intense phase-locked THz transients are generated by difference frequency mixing of two phase-correlated infrared pulse trains[15]. Figure 1a shows the waveform detected electro-optically (centre frequency: 30 THz, peak field in air: $E_a$ = 72 MV/cm = 0.7 V/Å). The field is focused into a bulk crystal of the semiconductor gallium selenide (GaSe), which forms a perfect host for coherent high-field transport. As the widths of valence and conduction bands are substantially smaller than the fundamental energy gap $E_g$, ultrafast carrier scattering by impact ionization[11] is suppressed. The pulse emitted from the sample differs sharply from the THz wave incident under an angle of θ = 70°. In the electro-optic signal recorded with a AgGaS$_2$ sensor (Fig. 1b), a high-frequency component at the second harmonic (SH) and a low-frequency offset are superimposed on the fundamental wave. The inset depicts the corresponding amplitude spectrum. Emission between 0.1 and 10 THz originates from optical rectification (OR), while the prominent peak at 60 THz stems from SH generation (see Supplementary Information for phase-matching considerations). The shape of the electro-optic waveform remains constant for consecutive scans, underpinning the absolute phase-stability of all frequency components.

Beyond second-order nonlinearities, ultrabroadband electro-optic traces also exhibit peaks at the third (90 THz) and fourth (120 THz) harmonics (Fig. 2a). Using a grating spectrometer with an InGaAs array detector, we record maxima at the 6$^{th}$ to 9$^{th}$ harmonics with decreasing intensity. A monochromator and a cooled silicon charge-coupled device monitor high-order harmonics (HH) with comparable intensities up to the 16$^{th}$ order (480 THz), which nearly coincides with the fundamental energy gap of GaSe at 476 THz. Interband photoluminescence attests to the presence of THz-induced electron-hole pairs. The existence of a plateau-like region is a hallmark of non-perturbative nonlinearities. HH generation (HHG) continues up to the 22$^{nd}$ order despite strong interband absorption. The total spectrum of 23 harmonic orders (including OR) covers 12.7 octaves from 0.1 to 675 THz (wavelength range: 3 mm to 440 nm), constituting a record bandwidth for tabletop THz sources. It is interesting to trace the HH intensity as a function of the THz amplitude, as shown e.g. for the 13$^{th}$ order in Fig. 2b. While the intensity initially scales asymptotically as $I_{13} \sim E_a^{26}$ (dashed line), a slower increase of $I_{13} \sim E_a$ (dotted line) observed at the highest fields confirms the non-perturbative nature of HHG. Other harmonic orders feature similar behaviour (Fig. S7).

Importantly, the entire THz HH spectrum is phase-locked. The CEP-stability of the low-frequency components has been established by electro-optic sampling (Fig. 1). Spectral interferometry of the HH comb with its own SH allows us to test the CEP stability above electro-optically accessible frequencies. Stable interference between the harmonic wave of order 2n and the SH of the n$^{th}$ harmonic is expected only if there is phase coherence between various orders. As an example, Fig. 2c represents the interferogram of the 12$^{th}$ and the frequency-doubled 6$^{th}$ harmonic. The fringe visibility of 81% proves excellent phase-coherence within the HH comb, while repeated scans (Fig. S2) highlight the long-term stability of the spectral phase with an rms jitter[28] of only 0.4 µrad. To the best of our knowledge, these results mark the first direct verification of CEP locking of HH radiation and point towards all-coherent electronic dynamics as its microscopic origin.

HHG in atoms has been explained semiclassically by a three-step model of consecutive ionization, acceleration, and recollision[30]. In contrast, THz acceleration of electrons in a periodic lattice potential is dominated by Bragg reflection due to the quantum mechanical wave nature of the electron. This dynamics is best described within a periodic band structure. Leadoff models of HHG in solids have captured the acceleration of pre-existing electrons in the conduction band[6-8], e.g. by integrating Bloch's acceleration theorem for a time-dependent external field. Yet this approach does not explain the creation of electronic population in the conduction band of an intrinsic semiconductor. For a consistent quantum description, we customize the theory of Ref. 10, accounting for both off-resonant excitation of interband polarization and intraband acceleration of electronic wave packets throughout the Brillouin zone (Fig. S3). A realistic representation of THz-induced dynamical band mixing is obtained by including three valence and two conduction bands (Fig. 2d). This approach incorporates the broken inversion symmetry of GaSe, required for the generation of even harmonic orders. Evaluating this model for the parameters of bulk GaSe[29], we obtain the spectrum shown in Fig. 2a (dashed line). The theory describes the occurrence of even and odd harmonic orders and yields good agreement with the experimentally observed relative intensities.

Our calculations identify dynamical Bloch oscillations combined with coherent interband excitation as the physical origin of the HHG process. Figure 3 shows the computed dynamics of electron wave packets for three different THz amplitudes. Electron-hole pairs are mainly created at negative field maxima (Figs. 3a and b) by coherent interband polarization. Simultaneously each half cycle of the carrier wave drives the electron momenta from zero up to their maximal value and back to zero. In the central part of the medium- to high-intensity pulses (Figs. 3c and d), electrons are sufficiently accelerated to reach the Brillouin zone edge and undergo more than one complete Bloch cycle within one half-cycle of the THz wave, corresponding to a Bloch period of only a few femtoseconds. This dynamics leads to high-frequency radiation (Fig. 2a) which is coherently locked to the driving field (Fig. 2c). The highest harmonic orders are predominantly caused by the coherent intraband current as confirmed by a switch-off analysis (Fig. S4), a semiclassical estimate (Fig. S9), and the dependence of HHG on the crystal orientation (Fig. S5).

Since the THz-induced coherent interband transitions involve several electronic bands, simultaneously, Bloch oscillations occur via a quantum interference of multiple excitation pathways (Fig. 2d). Therefore, the instantaneous phase and amplitude of the exciting THz field sensitively control the wave packet dynamics (Fig. S3) and we expect HHG to depend on the CEP $\varphi_{CEP}$ of the THz wave. Fig. 4 summarizes the measured and computed CEP dependence of the HH spectra. For $\varphi_{CEP} = 0.1\pi$, the HH spectrum is approximately sinusoidally modulated with a period of 30 THz, up to the bandgap (Fig. 4a). Beyond 476 THz the intensity decreases due to interband absorption. For its sign-flipped counterpart, i.e. $\varphi_{CEP} = 1.1\pi$, the shapes, the contrast, and the magnitudes of the HH peaks change, corroborating the interference between different excitation paths. For a comprehensive picture, we record (Fig. 4b) and compute (Fig. 4c) intensity spectra while varying $\varphi_{CEP}$ continuously from $-\pi$ to $4\pi$. HHG below 400 THz is only slightly affected by the CEP. In contrast the intensity maximum of the 16$^{th}$ harmonic shifts about the frequency of 480 THz, with a slope of -2.5 THz/rad. While the spectral shape repeats itself with a phase period of $2\pi$, the phase dependence is not sign-reversal invariant, in accordance with our quantum mechanical model.

In conclusion, ultraintense phase-locked THz pulses open a new chapter of extreme high-field transport in bulk semiconductors. Going beyond sub-cycle interband polarization in dielectric media[12,13], CEP-stable THz transients control quantum interference between coherent inter- and intraband transport, for the first time. This dynamics enables record-bandwidth CEP-stable spectra, continuously covering the entire THz-to-visible domain. Exciting perspectives in fundamental solid-state physics emanate from the fact that dynamical Bloch oscillations probe the electronic band structure of bulk media all-optically. Ultimately, ultrafast coherent carrier transport may inspire new routes towards future THz-rate electronics.

**Methods Summary**

**Experimental Setup**

The intense phase-locked multi-THz driving fields are extracted from a second-generation high-field source. A high-power Ti:sapphire laser amplifier (repetition rate: 3 kHz, pulse energy: 5.5 mJ, pulse duration: 33 fs) pumps two parallel optical parametric amplifiers. The spectrally-detuned signal pulse trains are mixed to generate inherently phase-locked multi-THz transients at the difference frequency (tuning range of centre frequency: 10 – 72 THz, pulse energy: up to 30 µJ). The CEP is controlled via the relative delay between the near-infrared generation pulses. This principle has been described in detail in Ref. 15. A synchronized 8-fs near-infrared pulse (centre wavelength: 840 nm, pulse energy: 10 nJ) serves as a gate for ultrabroadband electro-optic sampling. All experiments are performed at room temperature.

**Supplementary Information** is linked to the online version of the paper at www.nature.com/nphoton.

**Acknowledgments**

We thank Karl Renk for helpful discussions. This work has been supported by the European Research Council via Starting Grant QUANTUMsubCYCLE and the Deutsche Forschungsgemeinschaft (grant number KI 917/2-1).

**Author Contribution**

O.S., M.H., F.L., and R.H. conceived the study, O.S., M.H., F.L., B.U., C.L., and R.H. carried out the experiment, U.H., D.G., T.M., M.K., and S.W.K. developed the quantum mechanical model and carried out the computations, O.S., M.H., F.L., U.H., M.K., S.W.K and R.H. wrote the manuscript, all authors discussed the results.

**Author Information**

The authors declare no competing financial interests. Correspondence should be addressed to R.H. (rupert.huber@physik.uni-regensburg.de).

**Figure Legends**

**Figure 1 | Field-sensitive THz nonlinear optics. a,** The waveform of the THz driving field (blue, solid curve) features a Gaussian envelope (black, broken curve) with an intensity FWHM of 109 fs, which contains 3 optical cycles. The transient was recorded electro-optically in a GaSe sensor (thickness: 40 μm) with an 8-fs near-infrared gate pulse (centre wavelength: 0.84 μm). Inset: Corresponding amplitude spectrum. **b,** Electro-optic trace of the waveform generated by the intense THz pulse of panel **a** in a GaSe single crystal (thickness: 220 μm, angle of incidence: θ = 70°), kept at room temperature. The data are shown as recorded with an $AgGaS_2$ detector (thickness: 100 μm), not corrected for the detector response, which leads to a temporal walk-off between fundamental and SH. A superposition of the fundamental, SH and OR components is clearly seen. Insets: corresponding amplitude spectrum and experimental geometry, indicating the angle of incidence θ. The role of in-plane rotation of GaSe (angle φ) is investigated in Fig. S5.

**Figure 2 | CEP-stable THz high-harmonic generation in bulk GaSe**. **a,** HH intensity spectrum (solid line and shaded area) emitted from a GaSe single crystal (thickness: 220 μm, θ = 70°) driven by the phase-locked THz pulse in Fig. 1a. A combination of electro-optic sampling (EOS), an indium gallium arsenide diode array (InGaAs) and a silicon CCD (Si CCD) maps out the HH spectrum throughout the THz, far-infrared, mid-infrared, near-infrared, and visible regimes. $E_g/h$ marks the bandgap frequency (476 THz). The blue dashed curve shows the computed HH intensity spectrum, obtained from a five-band model (see Supplementary Information for details). **b,** Dependence of the intensity $I_{13}$ of the 13$^{th}$ harmonic on the incident THz amplitude $E_a$ (top scale). Bottom scale: internal THz amplitude, $E_{int}$, obtained from $E_a$ by accounting for reflection at the crystal surface. Spheres: experimental data, broken line: scaling law $I_{13} \sim E_a^{26}$, dotted line: scaling law $I_{13} \sim E_a$. **c,** The spectral interference between the frequency-doubled 6$^{th}$ harmonic and the 12$^{th}$ harmonic confirms CEP stability of the HH radiation. For experimental details of f-2f interferometry, see Fig. S2. **d,** Model electronic band structure of GaSe between Γ and K point, underlying the subsequent computations. Two conduction bands (CB1, CB2) and three valence bands (VB1 - VB3) are accounted for. Coherent excitation of electrons (spheres), e.g., from the second valence to the lowest conduction band can proceed via interfering pathways with different scaling in powers of the THz field.

**Figure 3 | Sub-cycle carrier dynamics driven by CEP-stable THz transients. a,** Phase-locked THz waveform driving coherent interband excitation and carrier transport. The calculated temporal dynamics of the distribution of conduction-band electrons, $n_e = f_k^{e1}$, is shown as color maps, for peak electric fields $E_{int}$ of **b,** 8 MV/cm, **c,** 11 MV/cm and **d,** 14 MV/cm. When the electron's wave vector, $k$, reaches the Brillouin zone boundary at the K-point, Bragg reflection occurs and the sign of $k$ is instantly inverted. White lines trace the centre of the electron distribution for different delay times.

**Figure 4 | CEP control of THz high-harmonic generation in GaSe**. **a,** HH intensity spectra measured for a CEP of the driving field of $\varphi_{CEP}$ = 0.1π (blue) and 1.1π (red). We define $\varphi_{CEP}$ to be zero when a positive field maximum of the carrier wave coincides with the peak of the Gaussian envelope. Inset: corresponding waveforms of the driving fields. **b**, Systematic dependence of HH spectra on the CEP of the THz transient. Broken lines highlight the spectral maxima of select harmonic orders. While HHG depends only moderately on the CEP for orders n < 15, CEP variation leads to a pronounced frequency shift for n ≥ 15. For n = 16, the frequency slope is -2.5 THz/rad. **c,** CEP dependence of the HH spectra computed via the quantum 5-band model.

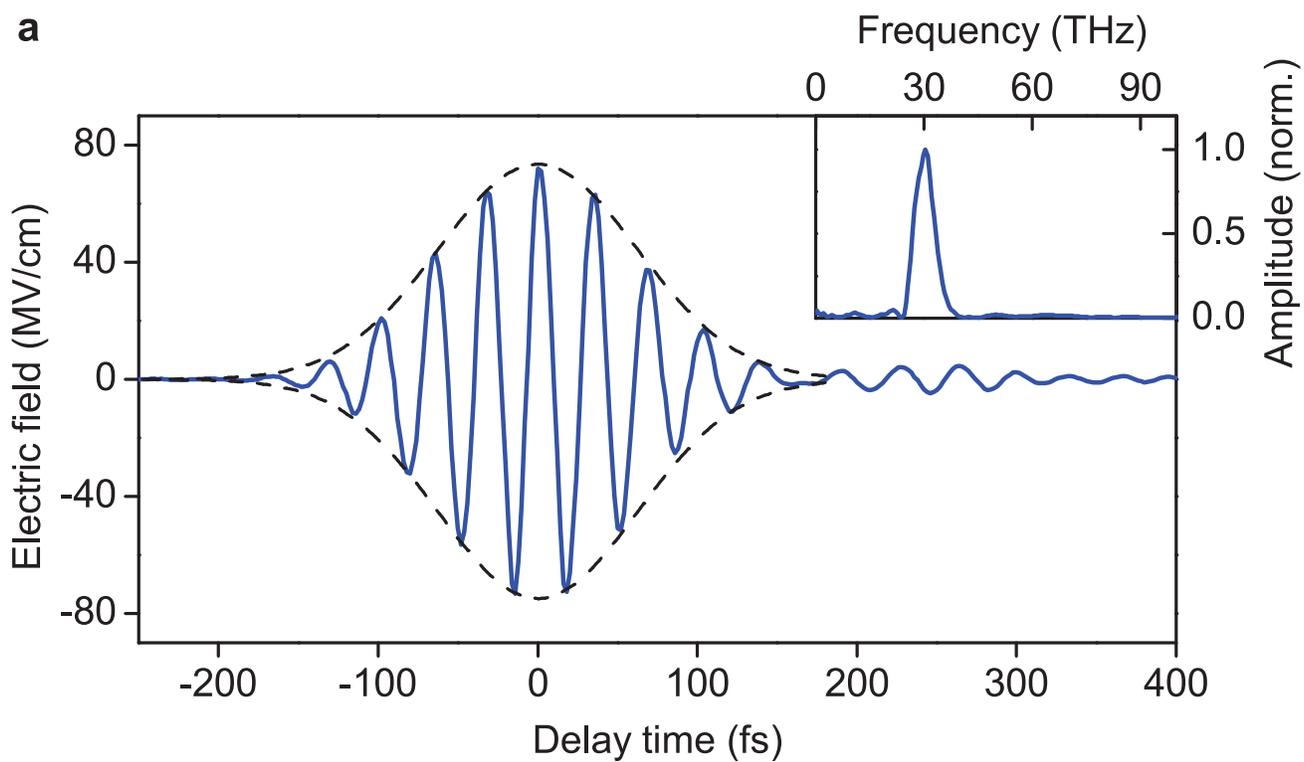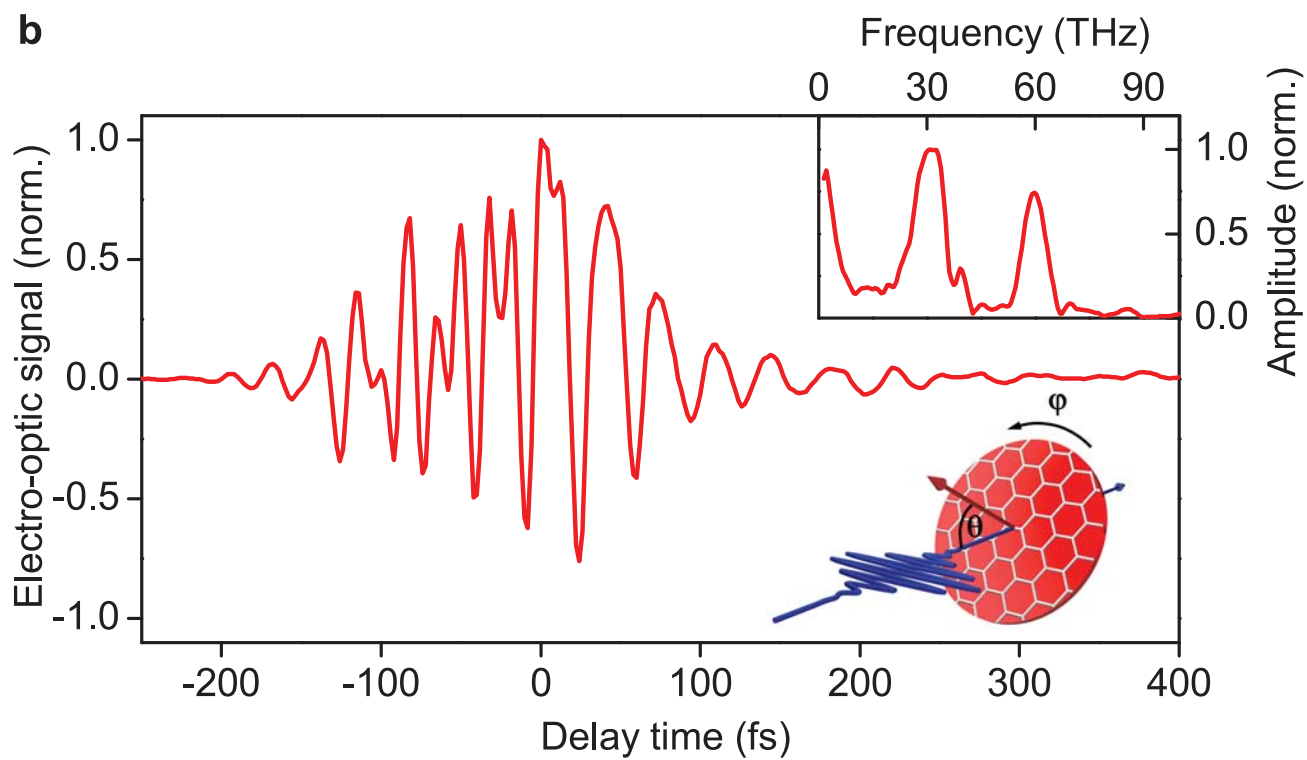

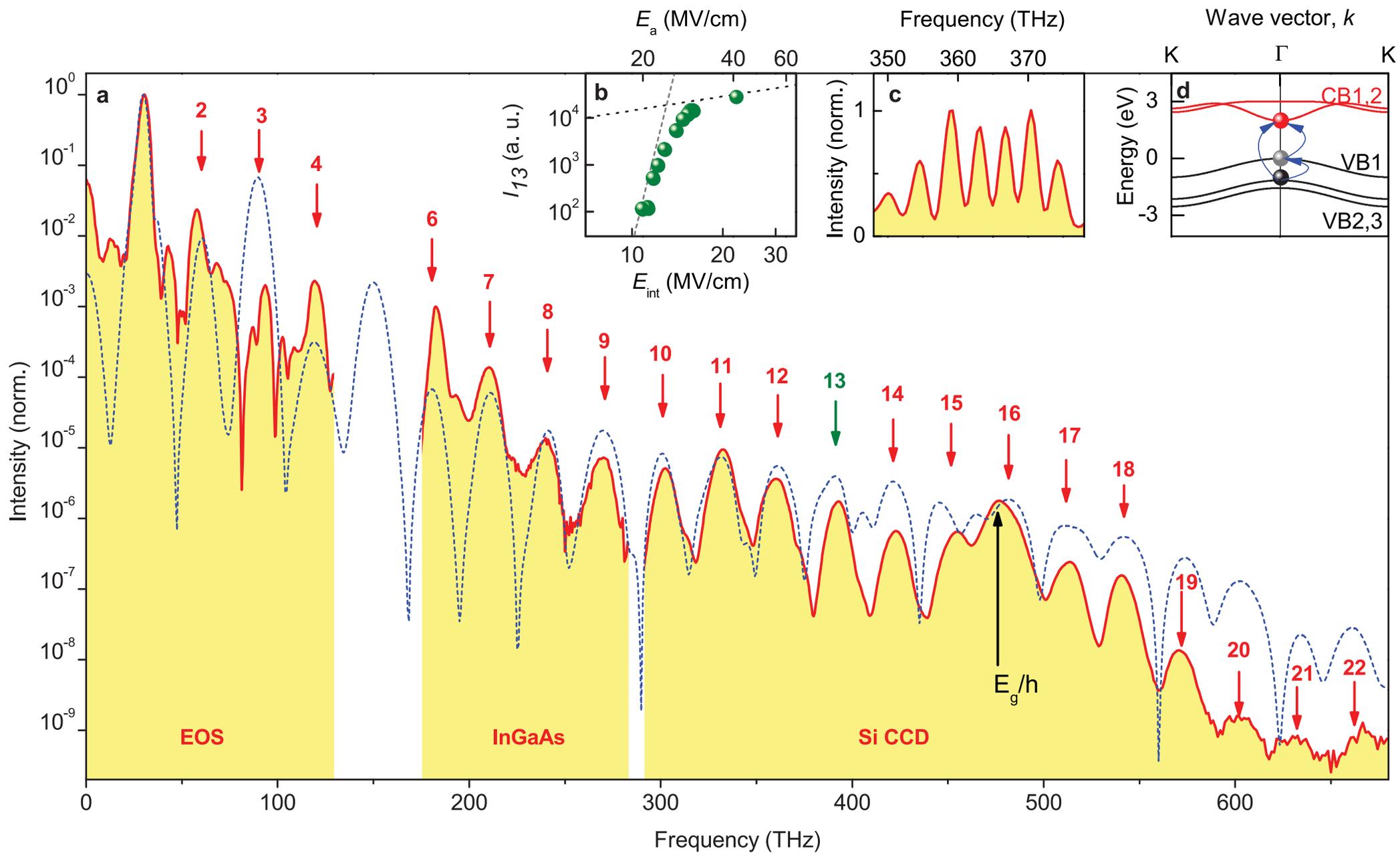

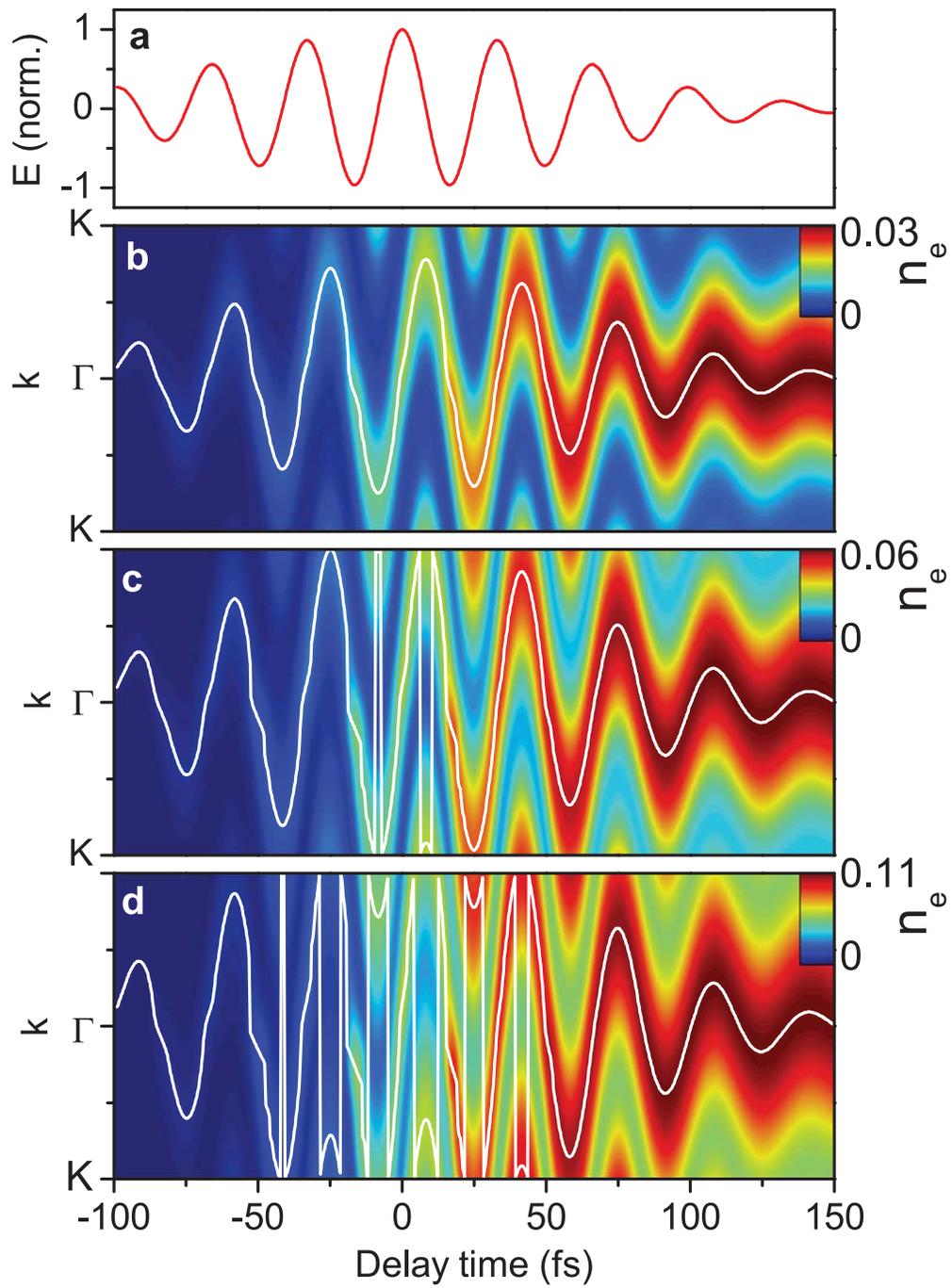

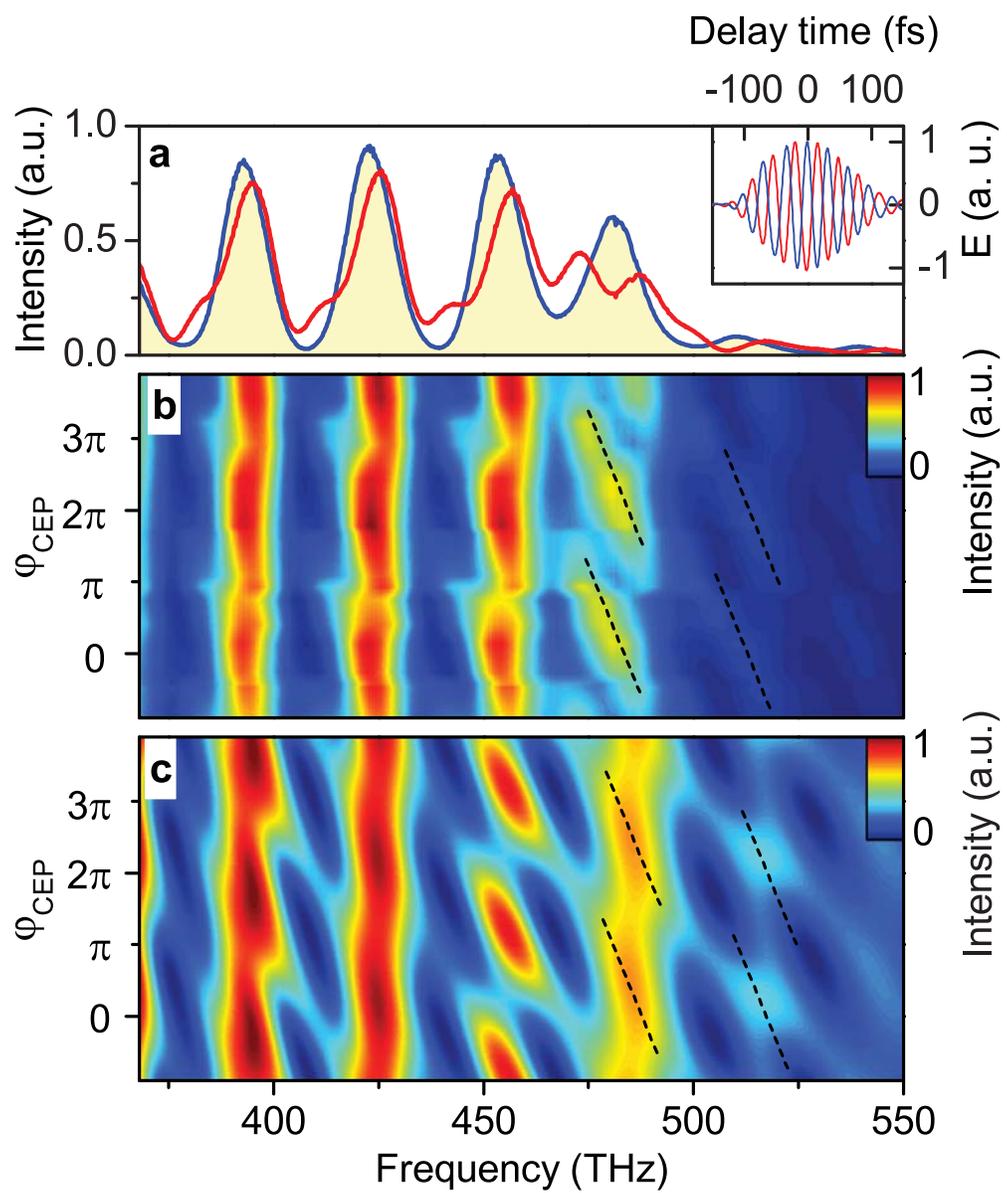